\documentclass[twocolumn]{article} 
\usepackage{epsfig}
 
\newcommand{\absosq}[1]{\bigl\vert#1\bigr\vert^2} 
\newcommand{\braket}[2]{\langle#1\vert#2\rangle} 
\newcommand{\ketbra}[2]{\vert#1\rangle\langle#2\vert} 
\newcommand{\ket}[1]{\vert#1\rangle}

\newcommand{\be}{\begin{equation}} 
\newcommand{\ee}{\end{equation}} 
\newcommand{\br}{{\bf r}}
\newcommand{\X}{$\cal X$}
\newcommand{\RRR}{$I\hspace{-0.25em}R^3$}
 
\begin{document}
%\raggedbottom
\title{QUANTUM MECHANICS\\
AND THE\\
COOKIE CUTTER PARADIGM} 
\author{Ulrich Mohrhoff\\ 
Sri Aurobindo International Centre of Education\\ 
Pondicherry-605002 India\\ 
\normalsize\tt ujm@auroville.org.in} 
\date{} 
\maketitle 
\begin{abstract}
\noindent What has so far prevented us from decrypting quantum mechanics is the Cookie Cutter 
Paradigm, according to which the world's synchronic multiplicity derives from surfaces that carve up 
space in the manner of three-dimensional cookie cutters. This insidious notion is shown to be rooted 
in our neurophysiological make-up. An effort is made to liberate the physical world from this innate 
fallacy.
\end{abstract} 
 
\section{\large INTRODUCTION} 

We live in two worlds, each with its own kind of space, time, form, and substance. There is the {\it 
phenomenal} world, extended in phenomenal space and changing in phenomenal time, containing 
forms that at bottom are bounded regions filled with qualia (introspectible properties like red or 
green). And then there is the {\it physical} world, whose spatial aspect is given by the spatial relations 
that obtain among material objects, whose temporal aspect is given by the temporal relations that 
obtain among actual events and/or states of affairs, and whose forms are sets of possessed spatial 
relations---as will be shown in this article.

The first lesson of science is that appearances are deceptive. Things are not what they seem. It is a 
commonplace of the contemporary scientific world view that matter is intrinsically particulate even 
though phenomenal space is perfectly homogeneous. It is another commonplace that molecules are 
not bounded by colored surfaces even though all objects in phenomenal space are. We are 
sufficiently aware that the concept of ``form'' appropriate for phenomenal objects (a bounding 
surface) and the concept of ``matter'' that goes with it (continuous stuff filling bounded regions of 
phenomenal space) are not applicable to physical objects (objects in the physical world). But we are 
not yet sufficiently aware that the concept of ``form'' appropriate for physical objects and the concept 
of ``matter'' that goes with it, entail a concept of ``space'' that is inconsistent with the standard 
mathematical description of physical space. This concept of ``space'', moreover, entails a concept of 
``time'' that is similarly incompatible with the standard mathematical description of physical time. 
Where physical time and space are concerned, we have yet to learn the principal lesson of 
science---appearances are deceptive.

This article serves three purposes. The first is to show that the seemingly insurmountable 
interpretational problems raised by quantum mechanics have their roots in ways of thinking about 
(and mathematically representing) space and time that are adequate for dealing with the phenomenal 
world, but that are as inconsistent with the ontological implications of quantum mechanics as the 
notion of absolute simultaneity is with special relativity. The second purpose is to arrive at appropriate 
ways of thinking about physical space and time and at adequate concepts of ``form'' and 
``substance''. The third purpose is to pinpoint the fallacy underlying the inappropriate concepts and 
ways of thinking---the Cookie Cutter Paradigm (CCP)---and to trace it to its neurophysiological roots. 
Because it has neurophysiological underpinnings and thus is, in a sense, innate, rejecting the CCP is 
no easy task. Yet it is worth the effort, for at the end we shall discover that the mathematical 
elegance and simplicity of quantum mechanics is matched by the depth and transparency of its 
ontological message.

Section~2 uses the paradigmatic two-slit experiment with electrons to argue that the reality of any set 
of spatial distinctions (i)~depends on the system and (ii)~is contingent on the existence of a matter of fact 
about the value of the corresponding position observable. What follows is that spatial 
distinctions cannot be regarded as inherent in space, and that, therefore, the standard, substantive, 
set-theoretic conception of space is the wrong conception of physical space.

Section~3 introduces a conception of space that is consistently applicable to the physical world. This 
portrays space as the totality of spatial relations existing among material objects. The corresponding 
concept of form is analyzed and shown to imply the formlessness of all fundamental particles. The 
popular belief that a fundamental particle has the form of a point is thereby shown to be an 
illegitimate importation from the phenomenal world (in which all objects have forms for psychological 
if not neurophysiological reasons) into the world of physics.

Section~4 discusses two properties of phenomenal space---the quality of continuous extension and 
the absence of intrinsic divisions---and clarifies their relation to physical space. Two ways of thinking 
about space and form, one based on how things appear in phenomenal space and one appropriate for 
quantum mechanics, are diagrammatically represented and compared.

Section~5 deals with the tricky question of how to think correctly about the simplest forms that exist in 
the physical world---those that consist of a single relative position. This involves two kinds of 
conditional probability assignments (one noncounterfactual and one counterfactual) and two 
corresponding kinds of probability (one subjective and one objective).

Objective probabilities betoken an objective indefiniteness, and this entails that the spatial and 
temporal differentiation of the physical world has a finite limit. This is discussed in Sec.~6, wherein 
one also finds definitions of the adjective ``macroscopic'' as related to positions, objects, clocks, and 
space.

Section~7 discusses another paradigmatic experiment, the scattering of particles at right angles, and 
arrives at the conclusion that, {\it intrinsically}, all fundamental particles are identical in the strong 
sense of {\it numerical} identity. What makes this conclusion {\it seem} preposterous is the insidious 
notion that prevents us from decrypting quantum mechanics---the CCP.

Section~8 examines various definitions of ``substance'' in respect of their applicability to (i)~a 
fundamental particle and (ii)~that which each fundamental particle intrinsically is. It is found that there 
is exactly one substance---pure, unqualified existence,---and that the physical world arises from it by 
the only logically consistent expedient---the realization of spatial relations. By entering into spatial 
relations with itself, existence acquires the aspect of a multiplicity of entities. But what exists at either 
end of each spatial relation is identically the same entity. All there is, at bottom, is existence and 
spatial relations between existence and itself.

The final section is devoted to the CCP---the notion that the world's synchronic multiplicity derives 
from surfaces that carve up space in the manner of three-dimensional cookie cutters. This is shown to 
be rooted in our neurophysiological make-up. Some of its consequences are examined to underline 
just how insidiously it prevents us from making sense of quantum mechanics.

\section{\large THE CONTINGENT REALITY OF\\
SPATIAL DISTINCTIONS} 

\label{distinctions}
The two-slit experiment with electrons requires no introduction~\cite{SeeFeynmanetal,Tonomura89}. 
According to Feynman~\cite{Feynmanetal65}, it ``has in it the heart of quantum mechanics''. But this 
appears to be something so baffling that we are warned off speculating about its significance: ``Do 
not keep saying to yourself, if you can possibly avoid it, `But how can it be like that?' because you 
will go `down the drain' into a blind alley from which nobody has yet escaped. Nobody knows how it 
can be like that''~\cite{Feynman67}. The last word on the subject apparently is that ``[e]lectrons seem 
to have modes of being, or modes of moving, available to them which are quite unlike what we know 
how to think about''~\cite{Albert92}.

As will be shown in this section, if ``[n]obody knows how it can be like that'', it is because we use the 
wrong concept of space. If the behavior of electrons in two-slit experiments is ``quite unlike what we 
know how to think about'', it is because we think about space as something that exists by 
itself---rather than by virtue of the spatial relations that obtain among the world's material 
constituents---and that contains, as subsets or partitions, all conceivable spatial 
divisions~\cite{Mohrhoff00,Mohrhoff}. 

If all conceivable spatial divisions were intrinsic to physical space, they would have an unconditional 
reality (that is, they would be real for every physical object), and one of the following statements 
would necessarily be true of every physical object $O$ contained in the union $R\cup R'$ of two 
disjoint spatial regions: $O$ is inside $R$; $O$ is inside $R'$; $O$ has two parts, one inside $R$ 
and one inside $R'$. If this were the case, no electron could ever pass through the union of two slits 
without passing through either slit in particular {\it and} without consisting of parts that pass through 
different slits. But this is precisely what electrons do when interference fringes are observed in 
two-slit experiments (and we do not postulate hidden variables).

What these interference fringes are trying to tell us is that spatial distinctions are not real {\it per se}. 
They have a conditional reality; they supervene---and thus depend---on the actual goings-on in the 
physical world. The conceptual distinction between two disjoint regions of space may 
be real for one physical object and nonexistent for another. If an electron can pass through the 
union of two slits without passing through either slit in particular and without consisting of parts that 
pass through different slits, it is because the distinction we make between the slits need not exist 
for the electron. The electron has not three {\it a priori} modes of going through the 
slits, as envisaged by Albert~\cite{Albert92}, but four:
\begin{tabbing}
(i) \hspace{0.7em} \= it takes the first slit and not the second slit;\\
(ii) \> it takes the second slit and not the first slit;\\
(iii) \> it takes the first slit and it takes the second slit;\\
(iv) \> it takes the ``undivided union'' of the slits.
\end{tabbing}
The first three modes assign separate truth values to the two statements ${\bf e}_1$ = ``the electron 
takes the first slit'' and ${\bf e}_2$ = ``the electron takes the second slit''. In so doing they entail that 
the distinction we mentally make between the regions defined by the slits, is real for the electron. 
(That no electron is ever found to move according to the third {\it a priori} mode betokens the 
electron's indivisibility. If we find an electron taking the first slit and we also find an electron taking 
the second slit, we observe two electrons, rather than one.) The fourth mode of moving, overlooked 
by Albert, is available to the electron if that distinction has no reality for the electron, or if nothing in 
the physical world corresponds to the difference between ${\bf e}_1$ and ${\bf e}_2$. Whereas (iii) is 
a conjunction of two propositions with inconsistent truth values, (iv) is a single proposition. The 
expression ``undivided union'' appears between quotation marks because, taken literally, it is 
self-contradictory: What is undivided cannot be a union. What the expression is intended to convey 
is that the division into separate regions of space, which exists in our minds, has no reality for the 
electron.

So when is this distinction real for the electron? The following is undisputed standard 
quantum mechanics: If something indicates the slit taken by the electron, or (equivalently) if there is a 
matter of fact about the slit taken by the electron, or (equivalently) if there is an actual event or state 
of affairs from which that slit can be inferred, then the electron has taken the indicated slit. In general: 
If something indicates the value taken by an observable $Q$ then $Q$ has taken the indicated value. 
The converse, too, is standard quantum mechanics, where ``standard'' connotes, in particular, the 
absence of hidden variables: If nothing indicates the slit taken by the electron, then the electron has 
not taken a particular slit. In general: If nothing indicates the value taken by an observable $Q$ then 
$Q$ has not taken any value whatsoever.

This (the converse) is not undisputed. There are ``quantum realists'' who deny the necessity of a 
property's being indicated and instead affirm that probability one is sufficient for the possession of a 
property~\cite{Redhead87}, or that an element of reality corresponding to an eigenvalue of an 
observable $Q$ exists at a time $t$ if the prior probability measure for the time $t$ has the pure form 
$\ketbra{\psi(t)}{\psi(t)}$ and $\ket{\psi(t)}$ is an eigen-``state'' of $Q$. But this is an error. As I have 
argued in detail~\cite{Mohrhoff00,Mohrhoff}, the contingent properties of quantum-mechanical 
systems (or the values of quantum-mechanical observables) are {\it extrinsic}: they supervene on 
what happens or is the case in the rest of the world. To paraphrase 
Wheeler~\cite{Wheeler,Wheeler83}, no property is a possessed property unless it is an indicated 
property~\cite{note1}.

If an observable $Q$ has $n$ eigenvalues $q_i$, a successful measurement of $Q$ leads to an 
actual event or state of affairs that indicates $n$ truth values, one for each proposition ``$Q$ has the 
value $q_i$'', $i=1,\ldots,n$. One of these truth values is ``true'', the remaining $n-1$ truth values are 
``false''. Any remotely feasible position measurement permits us to distinguish between at most a 
finite number of spatial regions (although for theoretical purposes we may allow a countable set of 
distinguishable regions). In order to fully characterize a position measurement---and the observable 
being measured---we must specify the regions between which the measurement can distinguish. 

As an illustration, let us consider a particle the support of whose wave function {\it for\/}---not {\it 
at\/}~\cite{note2}---the time $t$ is contained in $R$, and two arrays of detectors 
$\{D^1_i|i=1,\ldots,n\}$ and\linebreak
$\{D^2_k|k=1,\ldots,m\}$. The sensitive regions of each array of detectors 
are disjoint, and their union contains $R$. The two arrays thus measure different position observables 
on the particle, one capable of distinguishing between $n$ regions and assigning truth values to 
the $n$ propositions ``The particle is inside the sensitive region $R^1_i$ of $D^1_i$'', the other 
capable of distinguishing between $m$ regions and assigning truth values to $m$ propositions. 
If the first position observable is (successfully) measured, $n$ regions are distinct for the particle, 
and this finds expression in the existence of $n$ truth values for the $n$ propositions ``The particle is 
inside $R^1_i$''.

If position measurements are seen in this light, it becomes clear that the necessary and sufficient 
condition for attributing a value to a position observable $Q$ on a physical system $O$ is also the 
necessary and sufficient condition for the reality, for $O$, of the distinctions we make between the 
regions in the spectrum of $Q$. $Q$ has a value if and only if that value is indicated. But what is 
indicated is not just the value of $Q$. If $Q$ is capable of distinguishing between $n$ regions, $n$ 
truth values are indicated: For each of the $n$ regions, $O$'s presence in it or absence from it can be 
inferred. But the distinction we make between spatial regions has a reality for $O$ 
precisely if $O$'s presence in or absence from each region can be separately affirmed. Hence 
the distinction we make between the $n$ regions in the spectrum of $Q$ has a reality for $O$ if and 
only if the value of $Q$ is indicated or, equivalently, if and only if $Q$ has a value. Applied to the 
two-slit experiment with electrons this means that the distinction we make between the slits is real 
for an electron if and only if something indicates the slit taken by the electron, or (equivalently) if and 
only if the corresponding binary observable has a value, or (equivalently) if and only if both ${\bf 
e}_1$ and ${\bf e}_2$ possess truth values. If the propositions ${\bf e}_1$ and ${\bf e}_2$ are neither 
true nor false, they are meaningless, and the distinction we make between the slits has no reality for 
the electron.

The reality of spatial distinctions thus is contingent on what is, and what is not, indicated, and it also 
depends on the system. The standard, substantive, set-theoretic conception of space then cannot be 
the right conception of physical space. Physical space cannot be something that exists by itself, 
independently of the world's material constituents, and that by itself contains, as subsets, all 
conceivable spatial divisions, for if it were such a thing, the reality of spatial distinctions could not be 
contingent. All conceivable spatial distinctions would be real {\it per se}, and hence they would be 
real for all physical objects. The electron could not go through the ``undivided union'' of the slits, and 
no interference fringes could be observed.

\section{\large SPACE AND FORM IN THE\\
PHYSICAL WORLD}

\label{spaceform}
What, then, is the right way of thinking about physical space? One thing is clear: If space is not 
something that has intrinsic parts, the synchronic multiplicity of the physical world---the multiplicity of 
material objects~\cite{note3} that exist at any one time---cannot be defined in terms of the ``parts of 
space''. It can only be defined in terms of the material objects that exist at any one time and/or in 
terms of the spatial relations that exist among these objects.

The following is equally clear: If space is not something that exists by itself and has intrinsic parts, 
spatial relations cannot be attributed to its parts. Spatial relations exist to the extent that they are 
attributable to (pairs of) material objects. And since spatial relations is all it takes to specify the spatial 
aspect of the physical world, it is safe to affirm that physical space {\it is} the totality of spatial 
relations existing among the world's material constituents~\cite{note3a}. Here is another way of 
saying this: In order to describe the physical world, no reference to ``space'' is needed. All we need to 
refer to is the {\it positions} of material objects. And since these are relatively defined, reference to 
the relative positions of material objects---or, synonymously, to the spatial relations existing among 
material objects---is sufficient.

It follows that there is no such thing as an empty physical space. A physical world without material 
constituents is a spaceless world. It also follows that there is no such thing as a physical space 
containing a {\it single} material object, for two reasons. First, physical space contains spatial relations 
rather than objects. Second, a physical world containing a single material object lacks spatial 
relations; so it too is a spaceless world. (Such a world, moreover, would be indistinguishable from the 
single object it contains. It takes a minimum of two material objects to create a world containing 
something other than itself, such as spatially related objects. A world containing a single object is like 
the clapping of one hand.)

For the same reason that a physical world containing a single material object is spaceless, a 
noncomposite object---one that lacks parts---is formless. If the wrong way of thinking about physical 
space---as a substance in which all conceivable spatial divisions inhere---were the right way of 
thinking about physical space, the parts of a material object $O$ would be defined by surfaces 
partitioning the space ``occupied'' by $O$, and the form of $O$ would be defined by the surface 
bounding this space. A material object would have as many parts as the space it ``occupies'', and a 
material object without parts would have the form of a point. But there is no substantive space in 
which spatial divisions inhere. Therefore there are no surfaces by which the parts or the form of a 
material object could be defined. The ultimate parts of a material object are defined by their mutual 
spatial relations, and the form of a material object {\it is} the totality of its internal spatial 
relations---the spatial relations between its parts. A material object has as many internal spatial 
relations as it has pairs of parts. A noncomposite object lacks internal spatial relations, and therefore 
it also lacks a form.

While the form of a material object $O$ is the totality of its internal spatial relations, the position of 
$O$---considered not as an attribute but as an observable with contingently attributable 
values---essentially consists in $O$'s external spatial relations---those between $O$ or its material 
constituents and objects having no constituents in common with $O$. Two extreme cases are of 
special interest: (i)~If $O$ is the entire universe, there are no spatial relations external to $O$. Hence 
the universe as a whole lacks a position, which makes good sense. The form of the universe, on the 
other hand, being the set of its internal spatial relations, is identical with the set of all possessed 
spatial relations, which is space. (ii)~If $O$ is a fundamental (meaning, noncomposite) 
particle---according to the standard model of elementary particle physics, the quarks and the leptons 
are fundamental,---there are no spatial relations internal to $O$. Thus a fundamental particle has a 
position but it lacks a form.

This too makes good sense. Even on the inappropriate view of space as intrinsically and infinitely 
divided, the notion that a noncomposite object has a form would be redundant---as redundant as 
attributing positions to the ``points of space''. (The ``points of space'' are not objects that {\it have} 
positions. They {\it are} positions. They are properties that are available for attribution to material 
objects.) Saying that an object has the form of a point would be the same as saying that it has no 
parts, or that all its spatial relations are external, or that, relative to any given reference object, 
exactly one position can be attributed to it. There would be no reason to characterize a fundamental 
particle as having a pointlike form {\it over and above} being a noncomposite object.

The reasons for the popular belief that a particle has both a position and a (pointlike) form, have 
nothing to do with physics. They are psychological. They are to be found in the phenomenal 
world---the way we {\it perceive} the physical world. In the phenomenal world, the existence of spatial 
relations presupposes the existence of forms---visual percepts---to which spatial relations are 
attributable. Forms are {\it phenomenally} prior to spatial relations. In the physical world the converse 
is true: The form of a material object being the totality of its internal spatial relations, spatial relations 
are {\it ontologically} prior to forms. Again, in the phenomenal world positions are (or seem to 
be~\cite{note4}) realized individually, by objects with (ideally pointlike) forms. In the physical world 
positions are not realized by position-marking forms. They are realized by means of external spatial 
relations, and if the corresponding relata lack internal spatial relations, they are formless.

In the days when an atom was still widely thought of as a miniature solar system, Werner 
Heisenberg, if I remember right, argued that if atoms are to explain what the phenomenal world looks 
like, they cannot look like anything in the phenomenal world---an insight we have yet to assimilate in 
full. In the phenomenal world, in which objects are first of all bundles of qualia, every object 
necessarily has a form. The same is true of a world that is modeled after the phenomenal world, in 
which the shapes of things are bounding surfaces and matter is a continuous stuff filling the bounded 
regions. We smile at these simple-minded notions, yet the idea that a fundamental particle has a 
(pointlike) form is what becomes of these notions in the limiting case in which the bounding surface 
shrinks to a point. The space-filling stuff then metamorphoses into a material object that has both a 
position and a pointlike form.

\section{\large CONTINUITY AND DISCRETE-\\
NESS IN THE PHYSICAL WORLD}

If we look into {\it phenomenal} space---the only space we {\it can} ``look into'',---we perceive, first, a 
{\it quality} that goes beyond quantitative determinations and, second, a {\it unity} that goes beyond 
the unity of a set, defined by Cantor~\cite{Cantor1932} as ``a Many that allows itself to be thought of 
as a One''.

The quality to which I refer is the continuous extension of phenomenal space. This is as much a 
qualitative feature of our visual percepts as is the sensation of turquoise that we experience when 
looking at a tropical lagoon. Locutions such as ``relative position'', ``spatial relation'', or ``distance'' 
connote this qualitative feature as much as they connote quantitative determinability. Distances 
possess, in addition to their (more or less precise) values, this qualitative aspect, and nobody lacking 
our pre-conceptual grasp of phenomenal space is in a position to know it, anymore than Mary, 
confined from birth to a black-and-white room, was in a position to know color~\cite{Jackson1986}.

To discern the unity I have in mind, consider the visual image of a finite line segment $L$. Let us 
label its end points with real numbers $a$ and $b$. $L$ is in an obvious sense divisible into smaller 
segments, but nothing in its image warrants the notion that it is intrinsically multiple, let alone that it 
is a concatenation of point individuals with the cardinality of the real numbers. While between the 
real numbers $a$ and $b$ there exists a nondenumerable set $S$ of real numbers, between the 
points labeled $a$ and $b$ there exists a perfectly continuous and intrinsically undifferentiated line 
segment $L$. (Note that the preposition ``between'' is used here in two distinct senses: a spatial 
sense---``between the points labeled $a$ and $b$''---and a nonspatial sense that signifies ``greater 
than $a$ and less than $b$''.) $S$ possesses something that $L$ lacks, namely the multiplicity of a 
set of intrinsically distinct elements. And $L$ possesses something that $S$ lacks, namely the quality 
of continuous spatial extension and the unity that goes with it---the unity of something that exists in 
advance of divisions. To thinkers from Aristotle to Kant and Gauss it appeared self-evident that 
points on a line are extrinsic to the line, that they are added features not contained in it. They 
considered the line itself and, by implication, space itself is inherently undivided and as existing in an 
anterior relationship to limits and divisions. ``Space is essentially one'', Kant wrote~\cite{Kant1781}, 
``the manifold in it\ldots\ arises entirely from the introduction of limits.'' 

Unfortunately we have become so accustomed to conflating the continuity of phenomenal space with 
the discreteness of the set \RRR\ of triplets of real numbers that it is now almost impossible for us to 
tease them apart. We tend to visualize the reals as a continuous line without realizing that the very 
act of visualization introduces a qualitative element that is not warranted by the mathematical 
construction of the reals. Conversely, in our attempt to get a conceptual grip on physical space we 
seize on the reals as a set that appears to contain sufficiently many elements to ``fill'' a continuous 
line and thus to possess its continuity. We even deprive ourselves of words that are needed to 
distinguish between $S$ and $L$, as when we apply the adjective ``continuous'' to a {\it set}.

I will not quarrel over mathematical practice, which is justified by its results. Nobody is forbidden to 
call a self-adjoint operator ``elephant'' and a spectral decomposition ``trunk'', which makes it possible 
to prove a theorem according to which every elephant has a trunk. What is illegitimate is to create 
the impression that this theorem has something to do with biological pachyderms. By the same 
token, nobody is forbidden to call ``continuous'' a set that is cardinally equal to (or greater than) the real 
numbers. But it is illegitimate to identify this ``continuity'' with the continuity of a line in 
phenomenal space. It is therefore preferable to reserve the word ``continuous'' for the continuity that 
is a unique pre-theoretical feature of our visual percepts and images.

Continuity, so defined, is a feature of objects in phenomenal space---that is, of visual percepts,---and 
it is not a feature of any mathematical set. Is continuity a feature of objects in physical space? The 
answer is ``Yes'', or at least ``Why not?''. Physical space is the set of all possessed spatial relations, 
and to each of these relations we can consistently attribute the quality of undivided spatial continuity. 
Consider the distance ${\cal D}(AB)$ between two material objects $A$ and $B$. We tend to think of 
${\cal D}(AB)$ as a ``quantity'', and we tend to attribute to it the multiplicity of a segment of the ``real 
line''. That is, we tend to take it for granted that there are as many places between the two objects as 
there are real numbers between $0$ and $d(AB)$, the value of ${\cal D}(AB)$, assuming that ${\cal 
D}(AB)$ has a value. In reality there are as many places in the physical world as there are material 
objects. The places at which objects may be located (that is, the relative positions that may be 
attributed to them) do not exist unless objects are located there (that is, unless there are objects that 
possess them). There is no physical multiplicity that could qualify as inherent in a single spatial 
relation. There aren't any points or places between $A$ and $B$ unless other material objects are 
situated between $A$ and $B$. It takes a third object $C$ to introduce another two distances ${\cal 
D}(AC)$ and ${\cal D}(CB)$ such that $d(AB)=d(AC)+d(CB)$. The same equation does not hold 
among the three distances ${\cal D}(AB)$, ${\cal D}(AC)$ and ${\cal D}(CB)$. None of these 
distances is the sum of anything. ${\cal D}(AB)$ is not a quantity ``made up'' of quantities; it is a 
relation that possesses both the quality of undivided spatial extension and a value $d(AB)$. It 
interposes no places between $A$ and $B$. If anything interposes a location between $A$ and $B$, 
it is another material object $C$.

Thus Kant was wrong after all. What is ``essentially one'' is not space but {\it each spatial relation}. 
Being the totality of all possessed spatial relations, space has the multiplicity of the set of all pairs of 
material objects. Nor does ``the manifold'' in space, as he calls it, arise from the introduction of limits. 
Kant labored under the delusion of the CCP---the notion that the world's synchronic multiplicity 
derives from surfaces that carve up space like three-dimensional cookie cutters,---as we all do, 
thanks to our neurophysiological make-up (see Sec.~\ref{ccp}). In reality the synchronic multiplicity of 
the physical world is due to the existence of spatial relations.

\begin{figure}
\begin{center}
\epsfig{file=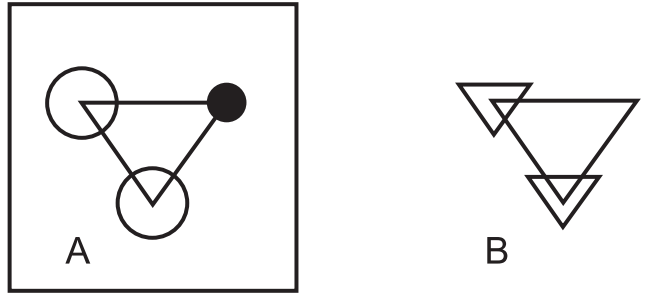,width=8.5cm}
\end{center}
\caption{Two ways of thinking about space and form. A---the naive view based on how things appear 
in phenomenal space. B---the correct view. The three lines forming the big triangle in both figures 
represent the spatial relations between three material objects. The big circles represent forms that 
are bounding surfaces. The small black circle represents a pointlike form. The small triangles 
represent forms consisting of spatial relations. The absence of a small triangle from one of the 
corners of the big triangle in B indicates the formlessness of a noncomposite object. The square 
frame represents a space that exists independently of material objects and ``contains'' them. It is 
absent from B because physical space is nothing more than the totality of all possessed spatial 
relations.}
\label{sf}
\end{figure}

Figure~\ref{sf} summarizes and contrasts two ways of thinking about space and form. The naive view, 
based on how things appear in phenomenal space, is illustrated by Fig.~\ref{sf}A, while the view 
appropriate for quantum mechanics is illustrated by Fig.~\ref{sf}B. The three lines forming the big 
triangle in both figures symbolically represent the spatial relations existing between three material 
objects. On the naive view, every material object has a form, which is either a bounding surface or a 
point. The big circles in Fig.~\ref{sf}A represent bounding surfaces; the small black circle represents 
a pointlike form. On the correct view, the form of an object is a set of internal spatial relations, 
and a noncomposite object is formless. The small triangles in Fig.~\ref{sf}B 
symbolically represent forms consisting of spatial relations. The absence of a small triangle from the 
third corner of the big triangle in Fig.~\ref{sf}B indicates the formlessness of a fundamental particle. 
The square frame, finally, represents a space that exists independently of material objects and 
``contains'' them. In Fig.~\ref{sf}B this frame is conspicuous by its absence. In the physical world, 
there is no space other than (i)~the set of all relative positions and (ii)~the quality of continuous 
spatial extension that is possessed by each relative position. While spatial relations in the 
phenomenal world derive their spatial character from phenomenal space, physical space derives its 
spatial character from the relations that it contains (in the proper, set-theoretic sense of 
``containment''). There is no physical space {\it over and above} the spatial relations that make up 
the forms of all beasts and baubles in the physical world.

\section{\large THE SHAPES OF SMALL THINGS}

\label{smallthings}
The non-visualizable character of the quantum world was recognized early on~\cite{Heisenberg49}. It 
led to the view that ``[t]he right way to understand quantum mechanics is not as a true description of 
physical reality but rather as an {\it instrument} for predicting the outcomes of laboratory 
experiments. There is no coherent interpretation of the quantum-mechanical formalism as describing 
an unobservable reality that is responsible for those experimental results. That reality is forever 
beyond our ken.''~\cite{Loewer98} What is ultimately responsible for the agnosticism of this view is 
our apparent inability to think of the spatial aspect of the physical world except in the terms laid down by the 
CCP, according to which (i)~the parts of a material object are defined by the parts of the space it 
``occupies'' and (ii)~the ``parts of space'' inhere in a space that exists independently of the objects it 
``contains''. This way of thinking about space may be appropriate for dealing with ordinary sensory 
perceptions but, as Bohr realized back in 1923, it is not appropriate for dealing with atoms: ``It is my 
personal opinion that these [interpretational] difficulties are of such a nature that they hardly allow us 
to hope that we shall be able, within the world of the atom, to carry through a description in space 
and time that corresponds to our ordinary sensory perceptions."~\cite{Bohr23}

Does the inadequacy of the standard, substantive, set-theoretic conception of space imply the 
impossibility of arriving at {\it any} adequate conception of the world's spatial aspect? If we subscribe 
to this {\it non sequitur} then we have to conclude with Stapp ``that `space,' like color, lies in the mind 
of the beholder''~\cite{Stapp72}. However, as the previous sections have shown, there {\it are} 
adequate ways of conceiving of the spatial aspect of the physical world. What ``lies in the mind of 
the beholder'' is the CCP and the concept of space it entails.

What is indeed strictly non-visualizable, is an individual fundamental particle: What does not have a 
form cannot be visualized. By no means, however, does this imply that we must resign ourselves to 
an agnostic attitude with respect to the real nature of a fundamental particle. Quite to the contrary, 
the formlessness of the fundamental constituents of matter tells us something of considerable 
ontological significance, for it implies, among other things, that fundamental particles exist in space {\it only} in the sense 
that they possess (more or less ``fuzzy'') relative positions. Considered in itself (that is, out of relation 
to other material objects), a fundamental particle has neither a position nor a form. It has no position 
because its position consists in its relations to other material objects; and it has no form because it 
lacks internal relations. But an object that has neither a position nor a form cannot be said to exist in 
space. Physical space exists {\it between} the fundamental particles; it is {\it spanned} by their spatial 
relations.

\begin{figure}
\begin{center}
\epsfig{file=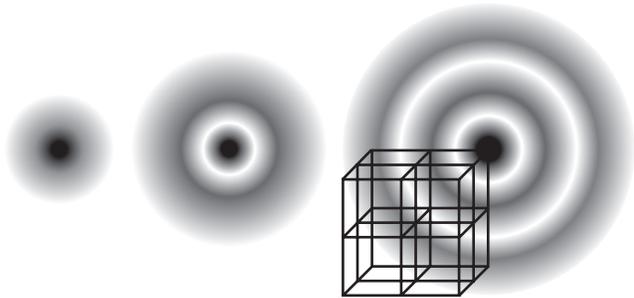,width=8.5cm}
\end{center}
\caption{The forms of a hydrogen atom in its first three spherically symmetric ``states''. Darker 
shades of gray represent higher ``probability densities'' (not to scale). Some nonexistent divisions of 
the third spherically symmetrical ``state'' are shown.}
\label{hydro}
\end{figure}

The simplest thing that we {\it can} visualize---provided that we are clear about the significance of the 
various features of our visualization---is a single relative position. As an example, let us visualize a 
hydrogen atom in one of its spherically symmetric stationary ``states'' (Fig.~\ref{hydro}). If we 
ignore the proton's internal relative positions (or if instead we visualize a positronium atom) then all 
there is to be visualized is the position of the electron relative to the other particle (or vice versa). 
Since this is the atom's only internal relative position, it constitutes its form. It is a most peculiar kind 
of form in that it has no parts. If it had parts, there would be as many positions relative to the other 
particle as there are parts, but there is only this one relative position.

Yet the cloudlike pictures in Fig.~\ref{hydro} do have parts, some of which are outlined for the third spherically symmetric 
``state''. What do these parts represent? And what do the varying densities of those clouds 
represent? One thing is clear: The parts are not features of any actual state of affairs; neither 
therefore are the local densities of the clouds. What the parts represent is something that is not the 
case but that would be the case if certain conditions were fulfilled. These are the conditions:

(i)~There exists an array of detectors $D_i$ with mutually disjoint sensitive regions $R_i$ such that 
$\bigcup_i R_i$ contains the support of the wave function $\psi(\br)$ associated with the electron's 
position relative to the proton.

A detector, here, is anything capable of indicating the electron's presence inside a particular region 
of the ``space'' \RRR\ of precise values associated with the electron's position relative to the proton 
(not to be confused with physical space).

(ii)~Exactly one of the detectors clicks, and thereby indicates the electron's presence in a particular 
region $R_k$.

If these conditions are fulfilled, the prior probability that the detector that clicks is the one with the 
sensitive region $R_k$, is given by
\[
p\,(R_k)=\int_{R_k}\!d^3r\,\absosq{\psi(\br)}.
\]
What is represented by the cloud, accordingly, is the ``probability density'' $\absosq{\psi(\br)}$. For 
two reasons this appellation is a misnomer. First, the probability that something happens in a 
particular region is not something that exists inside that region~\cite{note2}. Second, and more 
importantly, if conditions (i) and (ii) are not fulfilled, the distinctions we make between the regions 
$R_i$ have no reality for the atom's internal relative position, as was shown in 
Sec.~\ref{distinctions}. In this case there are no regions to which any kind of content (including 
``probability content'') can be attributed. {\it A fortiori} there are no infinitesimal regions to which an 
infinitesimal content $d^3r\,\absosq{\psi(\br)}$ can be attributed.

Figure~\ref{hydro} thus represents something that is {\it not} the case inasmuch as it represents {\it 
both} spatial divisions {\it and} probabilities that are distributed over them. If the spatial divisions are 
real for the electron (that is, if they are real with regard to its position relative to the proton), the 
probabilities $p\,(R_i)$ cannot be regarded as representing an objective feature of the physical world, 
for in this case the correct probability of finding the electron in a particular region $R_k$ is not 
$p\,(R_k)$ but either zero or one---the electron is inside one of these regions. (By ``correct'' I mean 
that the probability is assigned on the basis of all relevant facts, including the electron's presence in a 
particular region. $p\,(R_i)$ is our best guess as long as we are ignorant of the particular region 
containing the electron.) And if the probabilities $p\,(R_i)$ do represent an objective feature of the 
physical world, the conceptual partition of the ``space'' \RRR\ into the regions $R_i$ does not.

In what sense can probability assignments represent an objective feature of the physical world? The 
answer hinges on the fact that quantum-mechanical probability assignments, which are always 
conditional on the existence of a matter of fact concerning the value of a specific observable at a 
specific time, have two valid readings. The first is noncounterfactual: If the distinctions we make 
between the regions $R_i$ {\it are} real for the electron at a time $t$---the time at which one of the 
detectors clicks~\cite{note5},---$p\,(R_i)$ is {\it subjective} in that it is assigned without taking 
account of the particular region that contains the electron at the time $t$. It reflects our prior ignorance of 
that region.

The second reading is counterfactual: If (contrary to fact) the distinctions between the regions $R_i$ 
were real for the electron at a time $t$ (that is, if there were a matter of fact about the particular 
region containing the electron at the time $t$), the electron would be inside the region $R_i$ at the 
time $t$ with an objective probability $p\,(R_i)$. In this case $p\,(R_i)$ is {\it objective} in the sense 
that it is assigned on the basis of all relevant facts. No fact that has a bearing on the probability of 
finding the electron inside $R_i$ at the time $t$ has been ignored---there is nothing for us to be 
ignorant of~\cite{note6}.

The laws of quantum mechanics thus allow us to assign not only subjective probabilities to the possible 
results of performed measurements but also objective probabilities to the possible results of 
unperformed measurements. Objective probability assignments apprise us of an objective 
indefiniteness in the physical world, and they do so in the only adequate language---the language of 
counterfactuals.

It has been said that, although the terminology of ``indefinite values'' is prevalent in some elementary 
textbooks, what is really intended is that a certain observable does not possess a value at 
all~\cite{Redhead87a}. But there is more than that to the indefiniteness of the hydrogen atom's 
internal relative position, which finds expression in assignments of objective probabilities to 
counterfactuals. What essentially is implied by this indefiniteness is that the physical world is only 
finitely differentiated. As this has been discussed in detail in Refs.~\cite{Mohrhoff00} and 
\cite{Mohrhoff}, only a brief summary of the discussion will be given, for the sake of sketching a 
reasonably complete picture, in the following section.

\section{\large THE SPATIOTEMPORAL\\
DIFFERENTIATION OF THE\\
PHYSICAL WORLD}

Classical physics is consistent with the view that space is both intrinsically and infinitely 
differentiated. {\it Infinitely\/} because the position $q$ of a material object always has a precise 
value. This is the same as saying that for every partition $\{R_i\}$ of space into mutually disjoint 
regions, however small, and for every region $R_i$, the proposition ``$q$ is in $R_i$'' is either true or 
false. If we conceptually divide up space into infinitely many regions, there are infinitely many 
alternatives between which we can, in principle, distinguish. And {\it intrinsically} because this holds 
for {\it every} material object, which allows us to attribute the spatial divisions to ``space itself''.

In reality space is neither intrinsically nor infinitely differentiated. Physical space is the totality of 
relative positions existing between material objects, and {\it none} of these relative positions has a 
precise value. For every relative position we can find a partition $\{R_i\}$ of the ``space'' \RRR\ of 
precise values into sufficiently small but {\it finite} regions such that the corresponding position 
observable (cf. Sec.~\ref{distinctions}) {\it never} possesses a value. Thus one can never distinguish 
between more than a finite number of alternative regions, and this comes to saying that the physical 
world is only finitely differentiated spacewise.

However, while no relative position ever has a precise value, some relative positions are so ``sharp'' 
that their factually warranted values evolve in a completely predictable manner. Every indicated 
value (such as being inside region $R_k$) can be predicted, via the pertinent classical laws, on the 
basis of indicated earlier values. The indicated values of such relative positions evince no statistical 
variations; they are not {\it manifestly} fuzzy. It is therefore {\it quantitatively} correct (not only for all 
practical purposes but strictly) to treat these {\it macroscopic positions} as forming a self-contained 
system of causally connected, {\it intrinsic} positions---the classical domain---rather than a system of 
statistically correlated, {\it extrinsic} positions that presupposes position-indicating facts and, hence, 
the classical domain~\cite{note1}.

The sharp relative positions of classical physics ``mesh''; they can be embedded in a point set 
cardinally equal to \RRR, as spatial relations between points, and this point set can be identified with 
physical space. The fuzzy relative positions of quantum physics do not ``mesh''; they cannot be 
attributed to the members of a point set. Each relative position comes with its own ``space'' \RRR\ of 
precise values and potentially attributable regions, but the precise values and the attributable regions 
are both figments of our imagination. No actually existing detector has a sharply bounded (let alone 
pointlike) sensitive region.

Macroscopic positions lie somewhere in between. To take account of their fuzziness, they should not 
be embedded in a point set cardinally equal to \RRR. But they can be embedded, as spatial relations, 
in a {\it macroscopic space} whose 
elements are fuzzy ``points'' or fuzzily bounded ``regions''. Nothing needs to be said about the 
spacing of the ``points'' or the size of the ``regions'' except---

(i)~No finite portion of macroscopic space contains more than a finite number of such ``points'' or 
``regions''. This is warranted by the fact that one can never distinguish between more than a finite 
number of alternative regions, or the fact that physical space is only finitely differentiated spacewise.

(ii)~The ``points'' or elemental ``regions'' are sufficiently close or small to accommodate all 
macroscopic positions as relations between them.

The ``points'' or elemental ``regions'' being fuzzy, their relative positions are fuzzy. But if we let the 
``points'' or elemental ``regions'' be sufficiently small or close, we can make their relative positions 
sharper than the macroscopic positions ever are. This allows us to represent all macroscopic positions 
as spatial relations between the ``points'' or elemental ``regions'' of macroscopic space without loss of 
factually warranted precision. Macroscopic positions ``mesh'' in the sense that the relative positions 
of macroscopic objects can be attributed to the elements of macroscopic space. The elements of 
macroscopic space can then be attributed as positions to macroscopic objects. ({\it Macroscopic 
objects} are objects with macroscopic relative positions.)

Since macroscopic positions are not manifestly fuzzy, neither are the fuzzy ``points'' or fuzzily 
bounded ``regions'' of macroscopic space. Fuzziness implies probability distributions, and there is 
nothing over which the fuzzy points or boundaries could be probabilistically distributed. There is no 
finer partition than the partition of macroscopic space into its elemental ``regions''. Even this 
theoretical partition by definition exceeds the limit of the world's actual spatial differentiation. The 
probabilities corresponding to any finer partition are therefore objective and distributed over 
counterfactuals. Our conceptual spatial distinctions, carried beyond this limit, bottom out in a ``sea'' of 
objective probabilities.

The elements of macroscopic space are none of the things we are neurophysiologically disposed to 
attribute to space (see Sec.~\ref{ccp}). They are not mathematical points inasmuch as we can 
differentiate them, if only in our minds. They are not regions inasmuch as the sharpness of a 
region-defining boundary entails an unrealized degree of spatial differentiation. For the same reason 
they are not fuzzy ``points'' or fuzzily bounded ``regions''. They are ``regions'' only in the sense that 
we can differentiate them in our minds, and they are ``points'' only in the sense that they are not {\it 
actually} differentiated.

Many thinkers have been intrigued, and justifiably so, by the ability of the human mind to reproduce 
the physical world as faithfully as it does, or seems to do, depending on logic and mathematics 
alone~\cite{vW1980,Schilpp1949}. The success of physics is indeed astonishing, but the difficulties 
we experience in trying to understand quantum mechanics reveal that we are not all that 
well-equipped mentally. On the contrary, we seem to be neurophysiologically disposed to 
misunderstand quantum mechanics. The above attempt to characterize the {\it right} way of thinking 
about the actual extent of the world's spatial differentiation is a case in point. It starts from a {\it 
wrong} way of thinking about the spatial aspect of the world that appears to be innate, and then 
proceeds by elimination (points that are unlike mathematical points, regions that are unlike 
mathematical regions\ldots). If it is in fact innate then this is the best we can do.

What is true of the world's spatial aspect is equally true of its temporal aspect. There is no such thing 
as an intrinsically differentiated time. What is temporally differentiated is physical systems, and every 
physical system is temporally differentiated to the extent that it passes through distinct successive 
states, in the proper sense of ``state'' connoting properties indicated by facts. Not only the positions of 
things but also the times at which they are possessed are extrinsic. While attributable positions are 
physically defined by the not manifestly fuzzy sensitive regions of macroscopic detectors, attributable 
times are physically defined by the not manifestly fuzzy times indicated by macroscopic clocks. 
(Since the sensitive regions of macroscopic detectors and the times indicated by macroscopic clocks 
can be treated as intrinsic, no vicious regress is implied.)

The finite temporal differentiation of any finite physical system is a direct consequence of the world's 
finite spatial differentiation. Every physical system is temporally differentiated to the extent that it 
passes through distinct successive states, and no finite system passes through an infinite number of 
such states in a finite time span. During such a time span, a macroscopic clock, indicating time by 
means of some macroscopic position, can indicate no more than a finite number of distinct times, and 
therefore there exist no more than a finite number of such times. These times are both ``instants'' and 
``intervals'' in the same sense that the elements of macroscopic space are both ``points'' and 
``regions''. They are fuzzy or fuzzily bounded, but only in relation to an imaginary background that is 
more differentiated timewise than anything in the physical world.

\section{\large FUNDAMENTAL PARTICLES}

\label{fp}
\begin{figure}
\begin{center}
\epsfig{file=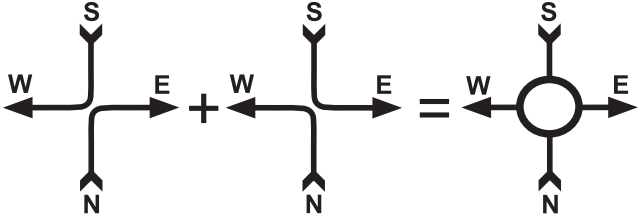,width=8.5cm}
\end{center}
\caption{Scattering of two particles at right angles. If the alternatives on the left-hand side of this 
symbolic equation are experimentally distinguishable, the equation holds for probabilities. In this 
case the diagram on the right-hand side gives an incomplete picture of what actually happens. If the 
alternatives are indistinguishable experimentally, the equation holds for amplitudes. In this case the 
diagram on the right-hand side gives the complete picture while the diagrams on the left-hand side 
are overspecific: The involve a distinction that Nature does not make.}
\label{scat}
\end{figure}

While the two-slit experiment with electrons can teach us how to think correctly about physical space 
and the shapes of material things, the following, equally paradigmatic (and equally enigmatic) 
scattering experiment can teach us how to think correctly about the substance aspect of the physical 
world. We will consider the special case in which there are two incoming particles, one ($N$) 
moving northwards and one ($S$) moving southwards, and two outgoing particles, one ($E$) moving 
eastwards and one ($W$) moving westwards. The assumed final state (one particle eastbound and 
one particle westbound) can be reached in either of two ways (Fig.~\ref{scat}).

It is a fundamental principle of quantum mechanics that the probability of a process $\cal P$ capable 
of following several alternatives depends on whether or not the alternative taken by the process is 
indicated or capable of being indicated. If something actually indicates the alternative taken, the 
probability of the process is given by the sum of the probabilities associated with its alternatives. By 
saying that the alternative taken by $\cal P$ is capable of being indicated I mean that the alternatives 
of $\cal P$ are correlated with the alternatives of another process ${\cal P}'$ such that a 
determination of the alternative taken by ${\cal P}'$ reveals the alternative taken by $\cal P$. In this 
case, too, the probability of $\cal P$ is given by the sum of the probabilities associated with its 
alternatives. Paradigm examples of such a situation are the experiments of Einstein, Podolsky, and 
Rosen~\cite{EPR,Bohm51} and of Englert, Scully, and Walther 
(ESW)~\cite{ESW91,ESW94,Mohrhoff99}. On the other hand, if nothing indicates or is capable of 
indicating the alternative taken by a process, the probability of the process is given by the absolute 
square of the sum of the amplitudes associated with its alternatives~\cite{note6a}.

Thus if there is a matter of fact about the slit taken by the electron, we add the probabilities 
associated with the alternatives ${\bf e}_1$ and ${\bf e}_2$, and if nothing indicates (or is capable of 
indicating) the slit taken by the electron, we add the amplitudes associated with these alternatives. 
By the same token, if there is a matter of fact about the alternative taken by the above scattering 
process, the probability of the process is given by
\[
p_e(E,W)=\absosq{\braket{EW}{NS}}+\absosq{\braket{WE}{NS}},
\]
and if nothing indicates the alternative taken, the probability of the process is given by
\[
p_i(E,W)=\absosq{\braket{EW}{NS}+\braket{WE}{NS}}.
\]
The subscripts $e$ and $i$ stand for ``exclusive'' and ``interfering'', respectively. $\braket{EW}{NS}$ 
and $\braket{WE}{NS}$ are the respective probability amplitudes associated with the alternatives 
($N\rightarrow E$, $S\rightarrow W$) and ($N\rightarrow W$, $S\rightarrow E$). For bosons 
$\braket{EW}{NS}=\braket{WE}{NS}$, so $p_b(E,W)$ is twice as large as $p_e(E,W)$:
\[
p_b(E,W)=\absosq{2\braket{EW}{NS}}=4\absosq{\braket{EW}{NS}}=2p_e(E,W).
\]
For fermions $\braket{EW}{NS}=-\braket{WE}{NS}$, so $p_f(E,W)=0$. Both results are inconsistent 
with the notion that the scattering process actually follows {\it either} of the alternatives shown in 
Fig.~\ref{scat}. If there is a matter of fact about the alternative taken (for instance, if the two particles 
are not of the same type and there is no possibility of type swapping), we are entitled to assume that 
the actual process follows either alternative, which leads to $p_e(E,W)$. If there isn't any matter of 
fact about the alternative taken, we cannot make this assumption, inasmuch it implies that the 
probability for scattering at right angles is given by $p_e(E,W)$, rather than by $p_b(E,W)$ or 
$p_f(E,W)$.

Once again we are confronted with a distinction that Nature does not make. If nothing indicates the 
alternative taken by the scattering process (which can only be the case if the two particles are of the 
same type), the distinction we make between the alternatives corresponds to nothing in the actual 
world. The process follows both alternatives, so we add amplitudes. Conversely, if something 
indicates the alternative taken by the scattering process (such as when the two particles are of 
distinct types), the distinction is real. The process follows either alternative, so we add probabilities.

The distinction we make between the two alternatives in Fig.~\ref{scat} is based on the assumption of 
{\it transtemporal identity\/}: We assume that the individual particles possess permanent identities. In 
those cases in which the distinction cannot be made (or if it is made, must be rescinded by adding 
amplitudes), that assumption cannot be made: The particles do not possess permanent identities. 
Hence they do not possess the property of ``being this very particle'', known to philosophers as 
``thisness'' or ``haecceity''. It follows that in those cases in which $p_e(E,W)$ is the correct 
probability, this is because the particles possess distinguishing characteristics and {\it not} because they 
possess something that would permanently individuate them even in the absence of distinguishing 
characteristics.

What, then, is the right way of thinking about two particles that lack distinguishing characteristics (and 
that therefore cannot be thought of as individuals that remain self-identical through change)? The two 
incoming particles as well as the two outgoing particles possess distinguishing 
characteristics: They travel in opposite directions, and they are in different places relative to the 
laboratory frame. But at the time of scattering no such distinguishing characteristics exist. How many 
things, then, exist at the time of scattering? According to the Identity of Indiscernibles, a principle of 
analytic ontology, there cannot be two absolutely indistinguishable things. Seen from the laboratory 
frame, the two particles are absolutely indistinguishable at the time of scattering, so on that account 
the answer should be: Only one thing exists at the time of scattering. Even at this time, however, the 
particles are in possession of a (more or less fuzzy) relative position, and this is sufficient for them to 
be two things even at the time of scattering.

And what is the right way of thinking about two indistinguishable particles {\it in themselves} (that is, 
out of {\it relation} to each other)? Since there is no such property as ``thisness'', the existence of a 
(more or less fuzzy) relative position is not only sufficient but also necessary for two indistinguishable 
particles to be two. Apart form their relative position, there is nothing that could make them two. 
Hence {\it intrinsically} (out of relation to each other) the ``two particles'' are {\it not} two; they are 
identical, and this not in the weak sense of exact similarity but in the strong sense of {\it numerical 
identity}.

A preposterous conclusion! What makes it {\it seem} preposterous, however, is, at bottom, nothing 
but the CCP, for this tricks us into believing that the parts of a material object are defined by the parts 
of the space it ``occupies''. If this were true of the physical world, no two physical objects could 
``occupy'' the same space. {\it A fortiori}, all physical objects would be intrinsically distinct for 
the same reason that, according to the CCP, all conceptually distinct regions of space are intrinsically 
distinct (see Sec.~\ref{ccp}).

\begin{figure}
\begin{center}
\epsfig{file=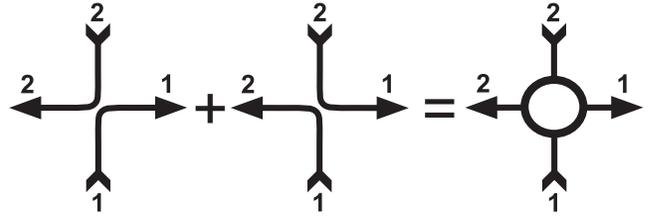,width=8.5cm}
\end{center}
\caption{Same as Fig.~\ref{scat}, except for the possibility of type swapping. Since the alternatives 
on the left-hand side (with and without type swapping) are indistinguishable experimentally, the 
diagram on the right-hand side gives the complete story, while the diagrams on the left-hand side are 
again overspecific; they involve a distinction Nature does not make.}
\label{swap}
\end{figure}

If type conversions are allowed, the conclusion that particles of the same type, considered in 
themselves, are numerically identical, can be extended to particles of different types. Suppose that 
the two particles in our scattering experiment are of different types (say, a proton and a neutron in, a 
proton and a neutron out), but that particles of the first type can be converted into particles of the 
second type (e.g., a proton into a neutron) and vice versa. Further suppose that $N$ and $E$ are of 
type~1, and that $S$ and $W$ are of type~2. Then the probability of this scattering event is given 
by
\[
p_i(E_1,W_2)=\absosq{\braket{E_1W_2}{N_1S_2}+\braket{W_2E_1}{N_1S_2}},
\]
where the indices specify the types to which the incoming and outgoing particles belong. Once again 
it is impossible to say whether a particular incoming particle is the same as or different from a 
particular outgoing particle, for the distinction that we make between the alternatives in 
Fig.~\ref{swap} is a distinction that Nature does not make. It exists solely in our minds.

The same argument that took us from $p_i(E,W)$ to the numerical identity of particles of the same 
type (considered in themselves), now takes us from $p\,(E_1,W_2)$ to the numerical identity of all 
particles of the same {\it basic} type (considered in themselves). What is characteristic of a basic 
particle species is that its members cannot be converted into members of a different basic species. 
How many basic particle species exist depends on the theory. According to the standard model, a 
member of one of the two species known respectively as hadrons and leptons cannot be converted 
into a member of the other species (the same applies to bosons and fermions), while in the so-called 
grand unified theories hadrons and leptons are mutually convertible, and in supersymmetric theories 
``once a fermion, always a fermion'' is no longer true either. In these theories, being a hadron, 
lepton, boson, or fermion is an accidental property of something that by itself is neither hadron nor 
lepton nor boson nor fermion; there exists just one basic particle species. But whether or not the final 
theory (assuming that there will be one) permits conversions between all particle types, the property 
of belonging to a particular type can be regarded as accidental or contingent, and all existing 
fundamental particles can be regarded as being {\it intrinsically} identical in the strong sense of 
numerical identity. What is more, they ought to be so regarded, for the parameters that characterize 
a particle species (mass, spin, and charges) tell us how a particle behaves {\it in relation to} other 
particles (specifically, how its external relative positions at different times are statistically correlated), 
but they tell us nothing about the particle's intrinsic behavior (that is, how it behaves {\it out of 
relation} to other particles).

\section{\large SUBSTANCE AND THE PHYSICAL\\
WORLD}

What, then, is this mysterious thing \X\ that all particles intrinsically are, and what is a particle, given 
its intrinsic numerical identity with every other particle? The least inappropriate concept is ``substance''. 
But this concept is laden with connotations that are inapplicable to the fundamental constituents of 
matter and must therefore be ``peeled off''. The only positive characterization we get from Aristotle, 
who introduced the concept, is that a substance remains self-identical through change, and this is 
precisely what cannot be said of a particle, although it may be said of that which all particles 
intrinsically are. For the rest Aristotle gives us negative definitions: Substance is neither predicable 
(``sayable'') of anything nor present in anything as an aspect or property of it. This too is something 
we may hold on to as a characterization of \X. As a characterization of a particle it won't do, for a 
particle may be said to be an instance or an aspect of \X.

For Locke, substance is that part of an individual thing in which its properties inhere. This concept of 
``substance'' is nothing but an imaginary hook from which the properties are supposed to 
hang~\cite{Russell}, a metaphysical glue that supposedly sticks them together. It may nevertheless 
be useful as a (partial) characterization of \X, for we may say that it is by virtue of \X\ that particles 
exist in the same world: \X\ is what ties them together as members of a single ontology~\cite{note7}. 
But it is definitely useless as an intrinsic characterization of a fundamental particle, for, in itself, this {\it 
has no properties that need to be tied together}. Recall that not only its contingent properties (its 
external relative positions, the corresponding relative momenta, and its spin components relative to 
any given axis) but also its dynamical parameters (mass, spin, and charges) tell us nothing but how 
the particle behaves {\it in relation to} the rest of the world.

Substance has been conceived again as that which is capable of existing independently of anything 
else. By this definition, too, \X\ can be considered a substance, while an individual particle cannot. 
The existence of each fundamental particle depends on that which it intrinsically is, and the existence 
of a multiplicity of fundamental particles depends on the existence of spatial relations.

Finally, there is the ordinary sense of ``substance''---what a thing is made of. This corresponds to 
Aristotle's understanding of individual things as composites of matter and form, form being a 
bounding surface, and matter being the extended stuff that fills it. In Sec.~\ref{spaceform} we have 
found the notion that a fundamental particle is a pointlike object to be a limiting 
case of this naive view: If the bounding surface shrinks to a point, what remains is a bit of stuff with a 
pointlike form. In reality there is no pointlike form; {\it a fortiori} there is no stuff that has one. What 
we {\it can} say of a fundamental particle is that it is ``made of'' \X, but only in the sense that 
intrinsically it {\it is} \X. In this sense, \X\ is the substance of every fundamental particle, and every 
material object is made of \X\ plus the spatial relations that constitute its form. This contrast sharply 
with the Aristotelian understanding that every material object is made of spatially extended stuff and a 
boundary that constitutes its form.

We get a better grip on the fundamental constituents of matter if we resolve the following apparent 
contradiction. (i)~Intrinsically (considered out of relation to other things), each fundamental particle 
is~\X. (ii)~Intrinsically, a fundamental particle lacks properties: Divested of its dynamical parameters 
and its contingent properties, it is nothing. How can a particle be both something (namely \X) and 
nothing? The contradiction is spurious, for statement~(i) is about an {\it existing} fundamental particle 
(an actual ingredient of the physical world), while statement~(ii) is about the {\it concept} of a 
fundamental particle (something that may or may not exist). By considering an existing particle out of 
relation to other things, statement~(i) divests this particle of its dynamical parameters and its 
contingent properties. Statement~(ii) further divests it of its {\it existence} as an actual ingredient of 
the physical world. The fact that in this case nothing remains, shows that the only thing that can be 
said of an existing fundamental particle, considered in itself, is that it exists. And this shows that all 
that \X\ bestows on an existing fundamental particle is {\it existence}.

If fundamental particles exist, there exist spatial relations, and if spatial relations exist, there exist 
material objects that either are fundamental particles or possess forms consisting of spatial relations 
between fundamental particles. We are inclined to think that the relations exist because the particles 
exist, but the opposite is at least equally true: The ultimate relata exist because the relations exist. 
The fundamental particles {\it qua instances} of \X\ owe their existence to the spatial relations that 
exist between them. Without these relations there would be no individual things; there would only be 
\X. The relations, on the other hand, owe their existence not to the relata but to that which they 
instantiate---to \X, which takes on the aspect of a multiplicity of relata because of the existence of 
relations.

I have just introduced a new concept: the (logical, rather than physical) process of {\it instantiation}. 
Instantiation is traditionally conceived as running parallel to predication: What gets instantiated is a 
predicable universal (an Aristotelian secondary substance or a Platonic Form like Horseness), and the 
resulting instance (e.g., a horse) is an impredicable individual. This way of thinking suggests that 
what is responsible for the instantiation is something that is present in the individual but absent from 
the universal, and this idea is at the root of the Platonic-Aristotelian dualism of Matter and Form and 
its subsequent transformations, including the erroneous idea that physical qualities are instantiated by 
the ``points of space''~\cite{Lewis86}.

The process of instantiation that takes us from \X\ to the fundamental particles of matter is something 
else altogether. As we have just concluded, all that \X\ bestows on an existing fundamental particle is 
{\it existence}. Hence all that gets instantiated by the spatial relations is existence. \X\ is existence 
pure and simple. Instead of arriving at this conclusion by analyzing experimental data, we could 
postulate pure existence and ask ourselves how it comes to be instantiated. To be effective, the 
instantiator must exist in advance of the instantiation. But in advance of the instantiation of existence 
there is nothing but existence. (If there existed anything else, it would be an instance of existence.) 
Hence only existence can instantiate existence. But existence can instantiate itself solely by entering 
into relations with itself, for this is the only process consistent with the proper logical dependencies: 
The instances of existence exist because the instantiating relations exist; the instantiating relations 
exist because \X\ enters into relations with itself; and \X\ exists because it is ``the one independent 
reality of which all things are an expression''---a dictionary definition of ``the 
absolute''~\cite{Audi1995}.

Thus if we adopt the premise of Parmenides that ultimately there exists a One 
Being~\cite{vWParm}---which is also the first assumption common to all his predecessors,---and if we 
avoid the pitfalls created by the CCP, we can see how the physical world arises from that One Being 
by the only logically consistent expedient---the realization of spatial relations. The relations account 
for the existence of material forms, and the intrinsic numerical identity of the relata  
accounts for the behavior of indistinguishable particles. This is 
sufficient reason to postulate something like an absolute or pure existence as a fundamental 
explanatory principle within physics itself. In good Spinozistic tradition (but with what seems to me a 
great deal more justification) I submit that there exists exactly one substance, that every fundamental 
particle, considered in itself, is this one substance, and that the particles, considered in relation to 
each other, are simply the relata that are logically entailed by the existence of spatial relations {\it 
between this one substance and itself}. By entering into spatial relations with itself, \X\ acquires the 
aspect of a multiplicity of entities, all of which are intrinsically as indeterminate as \X\ itself because 
intrinsically each of them is \X. What exists at either end of the spatial relation of any pair of 
fundamental particles is identically the same~\X. All there is, at bottom, is \X\ and spatial relations 
between \X\ and \X.

Spatial relations, as we have seen in Sec.~\ref{spaceform}, constitute both physical space and 
the shapes of material things. They are, moreover, {\it internal to} \X. Recall from 
Sec.~\ref{smallthings} that a fundamental particle, considered in itself, does not exist in space; 
instead, space is a web of relations spun {\it between} the fundamental particles. Add to this the 
conclusion that all fundamental particles are identically the same substance, and it follows that all 
spatial relations---and hence physical space itself---is internal to this one substance. The contrast 
with the vulgar conception of physical space as a container containing a multitude of self-existent 
substances (forms filled with stuff or pointlike bits of stuff) could hardly be greater.

According to Russell~\cite{Russell}, ``substance'' is a metaphysical mistake, due to transference to 
the world-structure of the structure of sentences composed of a subject and a predicate. Is this 
criticisms applicable to the one substance that I postulate as a fundamental explanatory principle? 
The answer is No, and the reason is that this postulate explains something that would otherwise 
remain unexplained. ``No acceptable explanation for the miraculous identity of particles of the same 
type has ever been put forward'', Misner et al.~\cite{Misneretal1973} wrote. ``That identity must be 
regarded, not as a triviality, but as a central mystery of physics.'' The ``miraculous identity of 
particles of the same type'' is exemplified by the scattering experiment discussed in Sec.~\ref{fp}. 
This is sufficiently explained by the {\it numerical} identity of all fundamental particles, considered in 
themselves. If we approach the experimental data with the assumption of transtemporal identity, as 
we are inclined to do, nothing can possibly explain the observed statistics of that experiment. But if 
we can bring ourselves to admit that identically the same substance exists at either end of each spatial 
relation, that statistics becomes a natural consequence of this numerical identity.

The real problem is not identity---be it the identity of this particle with that particle or the identity of 
this region here with that region there. The real problem is difference. Identity being the fundamental 
truth about the physical world, it is the proper starting point for all physical explanations. What needs 
explaining is not how two particles or two regions of space can be identical but how they become 
effectively distinct. No theory explains everything. Aristotle explained motion but did not explain rest. 
He considered it ``natural'' for bodies to be at rest. Newton explained acceleration but did not explain 
uniform motion. He considered it ``natural'' for bodies to be in uniform motion. Einstein explained 
geodesic deviation but did not explain geodesic motion. He considered it ``natural'' for bodies to move 
along timelike geodesics. Quantum mechanics explains geodesic motion in terms of interference---in the 
classical limit all trajectories but one interfere destructively---but 
leaves unexplained the nature and existence of quantum-mechanical interference. One day this too 
may be explained~\cite{boast}. But there is one question that no theory can aspire to answer, and this 
is the question of how is it that there is anything, rather than nothing. I would therefore consider truly 
fundamental only that theory which postulates nothing but pure, unqualified existence, and which can 
tell us how the physical world is both based on this existence (all particles are ``made of'' it) and 
suspended within it (space itself is internal to it). That theory is quantum mechanics. It tells us that the 
physical world arises from pure existence by a self-differentiation that involves nothing but the 
realization of spatial relations. These account for all the differences that we observe in the physical 
world at any one time---differences between locations and differences between things.

No physical theory can define what essentially distinguishes the actual world from a nomologically 
possible world (any world consistent with the laws of physics). {\it A fortiori}, no physical theory can 
account for the actuality of the actual world or the factuality of facts. This is something that classical 
physics did not have to worry about, for classical theories never dealt with the actual world. Classical 
physics concerned nomologically possible worlds, and the question as to which of these worlds 
describes the actual world was to be settled by observation rather than by any theory.

In quantum physics, 
by contrast, possibilities and facts are inextricably entwined. Quantum physics assigns probabilities 
to the possible results of possible measurements on the basis of the actual results of actual 
measurements. The theory talks about what {\it actually} happens or is the case. It distinguishes 
between (i)~alternatives all of which are mere {\it possibilities} (in which case it instructs us to add 
amplitudes) and (ii)~alternatives of which one is {\it factual} (in which case it instructs us to add 
probabilities). While classical physics is indifferent to the mystery of existence, quantum physics 
confronts us with it, and in two ways. For one, it presents us with property-indicating {\it facts} that are 
uncaused and therefore intrinsically inexplicable~\cite{Mohrhoff00}. In addition to that, it deals with 
fundamental particles, which in themselves are {\it existence} pure and simple. These numinous {\it 
apper\c cus} of ``bare reality'' play distinct ontological r\^oles. While the uncaused facts are the 
ultimate reason why there is something sayable, that which all particles intrinsically are is the ultimate 
reason why there is something of which anything can be said.

\section{\large THE COOKIE CUTTER\\
PARADIGM}

\label{ccp}
What ultimately prevents us from decrypting quantum mechanics (without extraneous 
additions like hidden variables~\cite{Bohm52} or spontaneous 
collapses~\cite{Ghirardietal86,Pearle89,Pearle97}, without using ``world'' in the plural~\cite{MWI73}, 
and without distinguishing between a mind-constructed ``internal'' or ``empirical'' reality and a 
mind-independent ``external'' or ``veiled'' reality~\cite{Putnam88,dEspagnat95} or bringing in 
consciousness in other 
ways~\cite{Albert92,Lockwood89,vN55,LB83,Peierls91,Page96,Stapp93,Mermin98}) is the 
CCP---the innate idea that the world's synchronic multiplicity derives from surfaces that divide up 
space like three-dimensional cookie cutters. The present section begins by explaining 
why this idea can be regarded as innate, and then focuses on how it prevents us from 
making sense of quantum mechanics.

We are adept at recognizing three-dimensional objects in drawings containing only outlines. (In fact, 
we can't help but perceive three-dimensional objects. We always see a Necker cube as pointing 
either in or out.) The reason why outlines are so easily recognized as objects lies in the manner in 
which the brain processes visual information.

The seminal work of Hubel and Wiesel~\cite{HW79} 
supports the following account. Visual information flows from retinal receptor cells via retinal 
ganglion cells to either of two lateral geniculate nuclei, and on to the primary visual cortex. The 
receptive field of each retinal ganglion or geniculate cell is divided into either an excitatory center 
and a concentric inhibitory surround (the ``on center'' configuration) or the reverse configuration (``off 
center''). (The group of retinal receptor cells from which a retinal ganglion or geniculate cell receives 
input is known as the cell's receptive field.) Thus an ``on center'' cell responds best to a circular spot 
of light of a specific size, responds well to a bright line that just covers the center (since then most of 
the surround is not covered by the line), and does not respond at all if both center and surround are 
fully and equally illuminated.

When visual information reaches the visual cortex, two major transformations take place. One leads 
to the fusion of input from both eyes, the other to a rearrangement of incoming information so that 
most of its cells respond to {\it specifically oriented line segments}. The optimal stimulus may be a 
bright line on a dark background or the reverse, or it may be a boundary between light and dark 
regions. One group of orientation-specific neurons responds best to lines with just the right tilt in a 
particular part of the visual field. Another group of neurons, receiving input from the first group, is 
less particular about the position of the line and responds best if the line is swept in a particular 
direction across the visual field.

These data suggest that visual representations arise by way of an analysis of the visual field that is 
based on {\it contrast information from boundaries between homogeneously lit regions}. Data arriving 
from homogeneously colored and evenly lit regions do not make it into conscious awareness. The 
interior of such a region is {\it filled in} on the basis of contrast information stemming from its {\it 
boundary}. This interpretation receives strong support from the remarkable faithfulness of color 
perception to the reflectances of colored surfaces, and its corresponding insensibility to the actual 
spectral composition of the radiances of such surfaces. (If color perception is based on discontinuous 
color changes across edges, continuous variations in illumination across the visual field go 
unperceived~\cite{Land77}.) It also explains why the blind spot is not perceived if it falls inside a 
homogeneous region (no sense data arrive from such a region anyway), and why outline drawings 
are so readily recognized as objects: The brain adds surfaces to outlines in the same way as it adds 
(unperceived) colored surfaces to (perceived) changes in color and brightness across edges.

While the visual field is inherently grainy (an array of retinal receptor cells), the phenomenal world is 
smooth. That we are unaware of the graininess is another consequence of how the brain processes 
visual information. Not only are uniformly colored regions of the visual field filled in homogeneously 
on the basis of contrast information across boundaries~\cite{note9}, but also the graininess of the 
boundaries is glossed over. Cells that respond to specifically oriented line segments in a particular 
part of the visual field receive input from cells with lined-up circular receptive fields. While the 
information coming from cells of the latter type is discrete, the perception of a line segment occurring 
when a cell of the former type is stimulated is continuous.

The phenomenal world thus is intrinsically a world of sharp and continuous boundaries filled with 
homogeneous content. Unlike the physical world, it {\it is} constructed from boundaries, and for this 
reason it conforms to the CCP~\cite{note8}. And if we naively believe that the physical world is 
constructed in the same way, we cannot but fail to make sense of quantum mechanics.

To see just how insidiously the CCP prevents us from decrypting quantum mechanics, let us examine 
some of its consequences. The idea that the synchronic multiplicity of the world depends on 
boundaries, or that the parts of a material object exist by virtue of the parts of the space it 
``occupies'', leads to the standard, substantive conception of an intrinsically and infinitely 
differentiated space. It is a substantive conception because it portraits space as intrinsically divided, 
and it does so because the parts of matter exist by virtue of the parts of space. (We cannot then hold 
the multiplicity of matter responsible for the multiplicity of space.)

Further, the parts of an intrinsically divided substantive space exist in themselves; each part---and in 
clear this means, each {\it conceivable} part---is a separate constituent of the world. But if {\it all} 
parts of space exist in themselves, there exist ultimate, not further divisible parts. (If each part were 
further divisible, we could not regard the division of space as completed. Hence we could not 
conceive of all parts of space as existing in themselves, independently of an always continuable 
process of division~\cite{note10}.) This is how the CCP tricks us into believing that space is 
adequately represented by (a point set cardinally equal to) \RRR.

Here are some additional consequences of the CCP:
\begin{itemize}
\item Every material object has a form, defined as the boundary of the space it ``occupies''.

\item A material object has as many parts as the space it ``occupies''. A noncomposite object 
therefore has the form of a point.

\item Two material objects cannot ``occupy'' the same space. Therefore all material objects are 
distinguishable by their positions.

\item Numerically sharp distances exist between all points of space. The distances between 
two material objects is the distance between two points of space. Hence numerically sharp distances 
exist between all material objects.
\end{itemize}
As the present article has demonstrated, none of these consequences of the CCP is consistent with 
what quantum mechanics is trying to tell us. Yet, to my knowledge, no interpretation of quantum 
mechanics except the Pondicherry Interpretation~\cite{Mohrhoff00} has ever challenged the CCP. 
This too goes to prove just how deep-seated a prejudice this fallacy is.


\begin{thebibliography}{99}%===============================
\bibitem{SeeFeynmanetal}
See for instance R.P. Feynman, R.B. Leighton, and M. Sands, {\em The Feynman Lectures in 
Physics}, Vol. 3 (Addison-Wesley, Reading, MA, 1965).

\bibitem{Tonomura89}
This classic thought experiment has been elegantly realized by A. Tonomura, J. Endo, T. Matsuda, T. 
Kawasaki, and H. Ezawa, ``Demonstration of single-electron buildup of an interference pattern'', Am. 
J. Phys. {\bf 57} (2), 117-120 (1989).

\bibitem{Feynmanetal65}
Reference~\cite{SeeFeynmanetal}, Sec.~1--1.

\bibitem{Feynman67}
Richard P. Feynman, {\it The Character of Physical Law} (MIT Press, Cambridge, MA, 1967), p.~129.

\bibitem{Albert92}
David Z. Albert, {\it Quantum Mechanics and Experience} (Harvard University Press, Cambridge, MA, 
1992), p.~11.

\bibitem{Mohrhoff00} 
Ulrich Mohrhoff, ``What quantum mechanics is trying to tell us'', Am. J. Phys. {\bf 68} (8), 728--745 
(2000); Eprint quant-ph/9903051.
%AJP: remove Eprint

\bibitem{Mohrhoff} 
Ulrich Mohrhoff, ``The One, the Many, and the Quantum'', Eprint quant-ph/0005110.
%AJP: add URL

\bibitem{Redhead87} 
Michael Redhead, {\it Incompleteness, Nonlocality and Realism} (Clarendon, Oxford, 1987), p.~72. 

\bibitem{Wheeler} 
``No elementary phenomenon is a phenomenon until it is a registered (observed) 
phenomenon.''---John Archibald Wheeler, ``Law without law'', in {\it Quantum Theory and 
Measurement} (Princeton University Press, Princeton, NJ, 1983), edited by J.A. Wheeler and W.H. 
Zurek, pp. 182--213.
 
\bibitem{Wheeler83} 
``No elementary quantum phenomenon is a phenomenon until it is registered, recorded, `brought to a 
close' by an `irreversible act of amplification,' such as the blackening of a grain of photographic 
emulsion or the triggering of a counter.''---John Archibald Wheeler, ``On recognizing `law without law' 
(Oersted Medal Response at the joint APS-AAPT Meeting, New York, 25 January 1983)'', Am. J. 
Phys. {\bf 51} (5), 398--404 (1983). It deserves mention that the characterization of a measurement on 
a system $S$ as involving an ``irreversible act of amplification'' is inadequate. Amplification achieves 
nothing but the entanglement of $S$ with a system $A$ possessing a large number of degrees of 
freedom: Instead of a probability measure over the possible properties of $S$, we then have a joint 
probability measure over the possible properties of both $S$ and $A$, which are statistically 
correlated. No amount of amplification can take us from a probability assignment to a 
property-indicating fact. On the other hand, once a property-indicating event or state of affairs has 
happened or come into existence, it is {\it logically} impossible to reverse this. For the relevant fact is 
not that the needle deflects to the left (which can be reversed by returning the needle to its neutral 
position); the relevant fact is that {\it at the time} $t$ the needle deflects (or points) to the left. This is 
a timeless truth. If at the time $t$ the needle deflects to the left, then it always has been and always 
will be true that at the time $t$ the needle deflects to the left.

\bibitem{note1} 
Quantum mechanics presupposes a classical domain encompassing property-indicating events or 
states of affairs, and therefore it can neither be inconsistent with nor account for the existence of this 
domain. How to think consistently about the coexistence of the classical an quantum domains has 
been explained in Refs.~\cite{Mohrhoff00} and \cite{Mohrhoff}.

\bibitem{note2} 
The functional dependence of a probability measure $\psi(\br,t)$ on $t$ is different from the time 
dependence of an actual state of affairs. A probability measure is not something that exists and 
evolves in time, anymore than it is something that exists in space. $\psi(\br,t)$ is ``prepared'' by 
a complete measurement, and the time to which it refers is the time of a subsequent measurement.

\bibitem{note3}
I use ``material object'' synonymously with ``physical object''.

\bibitem{note3a}
There is one conceivable spatial property that is not expressible in terms of the spatial relations that 
obtain in the physical world, namely the allegedly pointlike form of a fundamental particle. As the 
following will show, this pointlike form is not among the spatial properties of the physical world.

\bibitem{note4} 
Even in the phenomenal world, positions are {\it quantitatively} realized as relative positions. A single 
object can define ``here'' qualitatively, but it is insufficient for quantifying (attributing a numerical value 
to) ``here''. For this purpose we need to know how ``here'' is quantitatively related to ``elsewhere''.

\bibitem{Cantor1932}
Georg Cantor, {\em Gesammelte Abhandlungen} (Springer, Berlin, 1932), edited by A. Fraenkel and 
E. Zermelo, p.~204.

\bibitem{Jackson1986}
Frank Jackson, ``What Mary didn't know'', J. Philos. {\bf 83}, 291--295 (1986).

\bibitem{Kant1781}
Immanuel Kant, {\it Critique of Pure Reason}, first (German) edition, 1781, p.~25.

\bibitem{Heisenberg49}
E.g.: ``These atomic laws can be only imprecisely transposed into visual images of the atom, for 
Planck's quantum hypothesis on which these laws are based contains on principle a non-visualizable 
element.''---Werner Heisenberg, {\it Wandlungen in den Grundlagen der Naturwissenschaft} (Hirzel, 
Z\"urich, 1949), p.~47.

\bibitem{Loewer98}
Barry Loewer, ``Copenhagen versus Bohmian interpretations of quantum theory'', Brit. J. 
Phil. Sci. {\bf 49}, 317--328 (1998).

\bibitem{Bohr23}
Letter to H\o ffding, 22.9.1923, Bohr Scientific Correspondence, microfilm No.~3, p.~5; in John 
Honner, ``The transcendental philosophy of Niels Bohr'', Stud. Hist. Phil. Sci. {\bf 13} (1), 1--29 
(1982).

\bibitem{Stapp72}
Henry Pierce Stapp, ``The Copenhagen interpretation'', Am. J. Phys. {\bf 40} (8), 1098--1116 (1972).

\bibitem{note5} 
$t$ is defined as the time at which the distinctions we make between the regions $R_i$ are real for 
the electron. If these distinctions have no reality for the electron then neither has the time~$t$. A {\it 
particular} time $t$ is real for a quantum system only if it is the indicated time of possession of an 
indicated contingent property (such as being inside a particular region 
$R_k$), as is explained in Refs.~\cite{Mohrhoff00} and \cite{Mohrhoff}.

\bibitem{note6} 
As discussed in Ref.~\cite{Mohrhoff00}, the relevant facts may concern not only properties possessed 
before the time $t$ but also properties possessed after this time. $p\,(R_i)$ is objective only if we make the additional 
assumption that there is only one relevant fact, indicating the possession of a specific energy 
value $E_j$ prior to the time $t$.

\bibitem{Redhead87a} 
Reference \cite{Redhead87}, p.~48.

\bibitem{vW1980}
``Difficult though it be to imagine physics either completable or incompletable, it is perhaps even 
more difficult to imagine that physics should be possible at all\ldots. Why can the multiplicity of 
events be subjected to the consequences of a few simple postulates?''---Carl Friedrich von 
Weizs\"acker, {\em The Unity of Nature} (Farrar, Straus and Giroux, New York, 1980), pp.~174--175.

\bibitem{Schilpp1949}
``The most incomprehensible thing about the world is that it is comprehensible.''---Einstein in {\em 
Albert Einstein: Philosopher-Scientist} (Library of Living Philosophers, Evanston, IL, 1949), edited by 
P.A. Schilpp, p.~112.

\bibitem{EPR}
Albert Einstein, Boris Podolsky, and Nathan Rosen, ``Can quantum-mechanical description of 
physical reality be considered complete?,'' Phys. Rev. {\bf 47}, 777-780 (1935); reprinted in Wheeler 
and Zurek (Ref.~\cite{Wheeler}), pp. 138--141.

\bibitem{Bohm51}
David Bohm, {\it Quantum Theory} (Prentice Hall, Englewood Cliffs, NJ, 1951).

\bibitem{ESW91}
Marlan O. Scully, Berthold-Georg Englert, and Herbert Walther, ``Quantum optical tests of 
complementarity'', Nature {\bf 351} (6322), 111--116 (1991).

\bibitem{ESW94}
Berthold-Georg Englert, Marlan O. Scully, and Herbert Walther, ``The duality in matter and light'', 
Scientific American {\bf 271} (6), 56--61 (December 1994).

\bibitem{Mohrhoff99}
Ulrich Mohrhoff, ``Objectivity, retrocausation, and the experiment of Englert, Scully and Walther'', 
Am. J. Phys. {\bf 67} (4), 330--335 (1999).

\bibitem{note6a}
If a process is capable of following several alternatives from one set $\{{\cal A}^1_k\}$ of mutually 
contradictory alternatives, there is at least one other set $\{{\cal A}^2_k\}$ of such alternatives that is 
complementary to $\{{\cal A}^1_k\}$. In an ESW experiment (Refs. \cite{ESW91}--\cite{Mohrhoff99}), for 
instance, the experimenters may determine either the slit taken by an atom (by identifying the 
microwave cavity containing the photon left behind by the atom) or the phase relation with which the 
atom emerged from the ``undivided union'' of the slits (by opening the electro-optic shutters and 
watching the photosensor situated between the cavities). The pair of slits and the pair of possible 
phase relations are complementary alternatives. If the experimenters determine neither the slit taken 
nor the phase relation, then not only the conceptual distinctions we make between the alternatives of 
each set correspond to nothing in the physical world, but also the conceptual distinction we make 
between the two sets of alternatives has no counterpart in the physical world. In this case we cannot 
add the amplitudes associated with the first set of alternatives (the atom takes this or that slit), for 
this would imply that the atom went through both slits with a definite phase relation, or that the 
conceptual distinction we make between the alternatives of the second set corresponds to something in the 
physical world.

\bibitem{Russell}
Bertrand Russell, {\it A History of Western Philosophy} (Simon and Schuster, New York, NY, 1945), 
pp. 201--202.

\bibitem{note7}
It might be held that it is their spatial relations that tie particles together as members of a single 
ontology; without those relations, they would not exist in the same sense or in the same world. In 
truth, as we have seen in Sec.~\ref{fp}, the particles would not be {\it many} without those relations; 
they would be altogether what they intrinsically are---identically the same thing \X. In the absence of 
spatial relations there is nothing that could be tied together.

\bibitem{Lewis86}
``[A]ll there is to the world is a vast mosaic of local matters of 
particular fact, just one little thing and then another\ldots. We have geometry: 
a system of external relations of spatiotemporal distance between points\ldots. 
And at those points we have local qualities: perfectly natural intrinsic properties 
which need nothing bigger than a point at which to be instantiated\ldots. And 
that is all\ldots. All else supervenes on that.''---David K. Lewis, {\it Philosophical Papers}, Vol. II 
(Oxford University Press, New York, NY, 1986), p.~X. 

\bibitem{Audi1995}
Robert Audi, {\it The Cambridge Dictionary of Philosophy} (Cambridge University Press, Cambridge, 
1995).

\bibitem{vWParm}
Carl Friedrich von Weizs\"acker, ``Parmenides and quantum theory'', in Ref.~\cite{vW1980}, 
pp.~379--400.

\bibitem{Misneretal1973}
C.W. Misner, K.S. Thorne, and J.A. Wheeler, {\it Gravitation} (W.H. Freeman and Company, San 
Francisco, 1973).

\bibitem{boast}
I intend to do this in another article.

\bibitem{Bohm52}
David Bohm, ``A suggested interpretation of quantum theory in terms of hidden variables'', Phys. 
Rev. 
{\bf 85}, 166--193 (1952); reprinted in Wheeler and Zurek (Ref.~\cite{Wheeler}), pp. 369--396.

\bibitem{Ghirardietal86}
G.C. Ghirardi, A. Rimini, and T. Weber, ``Unified dynamics for microscopic and macroscopic 
systems'', Phys. Rev. D {\bf 34}, 470--491 (1986).

\bibitem{Pearle89}
Philip Pearle, ``Combining stochastic dynamical state-vector reduction with spontaneous 
localization'', Phys. Rev. A {\bf 39}, 2277--2289 (1989).

\bibitem{Pearle97}
Philip Pearle, ``True collapse and false collapse'', in {\it Quantum Classical Correspondence} 
(International Press, Cambridge, MA, 1997), edited by Da Hsuan Feng and Bei Lok Hu, pp. 51--68.

\bibitem{MWI73}
Bryce S. DeWitt and Neill Graham, eds., {\it The Many-Worlds Interpretation of Quantum Mechanics} 
(Princeton University Press, Princeton, NJ, 1973).

\bibitem{Putnam88}
Hilary Putnam, {\it Representation and Reality} (MIT Press, Cambridge, MA, 1988), p.~113.

\bibitem{dEspagnat95}
Bernard d'Espagnat, {\it Veiled Reality} (Addison-Wesley, Reading, MA, 1995).

\bibitem{Lockwood89}
Michael Lockwood, {\it Mind, Brain and the Quantum} (Basil Blackwell, Oxford, 1989).

\bibitem{vN55}
John von Neumann, {\it Mathematical Foundations of Quantum Mechanics} (Princeton University 
Press, Princeton, NJ, 1955).

\bibitem{LB83}
Fritz London and Edmond Bauer, ``The theory of observation in quantum mechanics'', in Wheeler 
and Zurek (Ref.~\cite{Wheeler}), pp. 217--259.

\bibitem{Peierls91}
Rudolf Peierls, ``In defence of `measurement'$\,$'', Physics World {\bf 4} (1), 19--20 (1991).

\bibitem{Page96}
Don N. Page, ``Sensible quantum mechanics: Are probabilities only in the mind?'', Int. J. Mod. Phys. 
D {\bf 5}, 583--596 (1996).

\bibitem{Stapp93}
Henry Pierce Stapp, {\it Mind, Matter, and Quantum Mechanics} (Springer, Berlin, 1993).

\bibitem{Mermin98}
N. David Mermin, ``What is quantum mechanics trying to tell us?'', Am. J. Phys. {\bf 66} (9), 753--767 
(1998).

\bibitem{HW79}
D.H. Hubel and T.N. Wiesel, ``Brain mechanisms of vision'', Sci. Am. {\bf 241} (3), 150--162 (1979).

\bibitem{Land77}
Edwin H. Land, ``The Retinex theory of color vision'', Sci. Am. {\bf 237} (6), 108--128 (1977).

\vfill\pagebreak
\bibitem{note9}
Since this involves the transition from objective brain mechanisms to subjective visual percepts, just 
how the filling in is accomplished is presently as impenetrable as the question of how anything 
material can be conscious in the first place.

\bibitem{note8}
Recently Jos\'e Luis Berm\'udez ({\it The Paradox of Self-Consciousness}, MIT Press, Cambridge, 
MA, 1998) has marshaled a remarkable amount of evidence supporting the existence of a 
prelinguistic, non-conceptual, representational content of consciousness. On Berm\'udez's reading 
the evidence suggests that the prelinguistic infant perceives bounded segments of the world that are 
unified and coherent, that have certain properties (for instance, shape), and that stand in certain 
relations with each other. This too goes a long way towards establishing the innateness of the CCP.

\bibitem{note10}
This touches on the age-old conflict between the intellectual demand for completeness and the 
perception of continuity, which argues against it. Science being driven by the desire to know how 
things are in themselves, it tends to come down in favor of completeness. Quantum mechanics 
teaches us how these apparently conflicting demands are, in fact, consistent. What makes them 
seem inconsistent is again---what else could it be?---the CCP. 

\end{thebibliography}
\end{document}